\documentstyle[prd,aps,eqsecnum,floats,epsfig]{revtex}
\tighten
\begin{document}
%
 \twocolumn[
 \hsize\textwidth\columnwidth\hsize\csname@twocolumnfalse%
 \endcsname
%
\title{Constructing Hybrid Baryons with Flux Tubes}
%
%
\author{Simon~Capstick}
\address{
	Department of Physics 
	\& Supercomputer Computations Research Institute,\\
 	Florida State University, Tallahassee, FL 32306, USA\\
}
\author{Philip~R.~Page}
\address{
	Theoretical Division, Los Alamos National Laboratory,
	P.O. Box 1663, Los Alamos, NM 87545, USA\\
}
\date{\today}
\maketitle
\begin{abstract}
%
Hybrid baryon states are described in quark potential models as having
explicit excitation of the gluon degrees of freedom. Such states are
described in a model motivated by the strong coupling limit of
Hamiltonian lattice gauge theory, where three flux tubes meeting at
a junction play the role of the glue. The adiabatic approximation for
the quark motion is used, and the flux tubes and junction are modeled
by beads which are attracted to each other and the quarks by a linear
potential, and vibrate in various string modes. Quantum numbers and
estimates of the energies of the lightest hybrid baryons are provided.
\end{abstract}
 ]
\def\slash#1{#1 \hskip -0.5em / }  
\def\rmb#1{{\bf #1}}
\def\lpmb#1{\mbox{\boldmath $#1$}}
\def\nn{\nonumber}
\def\>{\rangle}
\def\<{\langle}
\newcommand{\Eqs}[1]{Eqs.~(\protect\ref{#1})}
\newcommand{\Eq}[1]{Eq.~(\protect\ref{#1})}
\newcommand{\Fig}[1]{Fig.~\protect\ref{#1}}
\newcommand{\Figs}[1]{Figs.~\protect\ref{#1}}
\newcommand{\Sec}[1]{Sec.~\protect\ref{#1}}
\newcommand{\Ref}[1]{Ref.~\protect\cite{#1}}
\newcommand{\Refs}[1]{Refs.~\protect\cite{#1}}
\newcommand{\Tab}[1]{Table~\protect\ref{#1}}
\renewcommand{\-}{\!-\!}
\renewcommand{\+}{\!+\!}
\newcommand{\sfrac}[2]{\mbox{$\textstyle \frac{#1}{#2}$}}
\def\etahat{\mbox{\boldmath $\hat{\eta}$}}
\def\beq{\begin{equation}}
\def\eeq{\end{equation}}
%
\sloppy
Experiments at new electron-scattering facilities, such as those in
Hall B at TJNAF, are expected to produce many new baryon states,
including those which are described in quark potential models as
having explicit excitation of the gluon degrees of freedom. Low-lying
baryon states present in analyses of $\pi N$ elastic and inelastic
scattering, such as the $P_{11}$ Roper resonance $N(1440)$ which has
the same quantum numbers as the nucleon, have been
proposed~\cite{Roper,QCDsr} as hybrid candidates. This is based on
extensions of the MIT bag model~\cite{bag} to states where a
constituent gluon in the lowest energy transverse electric mode
combines with three quarks in a color octet state to form a colorless
state, and on a calculation using QCD sum rules~\cite{QCDsr}. Hybrid
baryons have also been constructed recently in the large-$N_c$ limit
of QCD~\cite{yan}.

With the assumption that the quarks are in an S-wave spatial ground
state, and considering the mixed exchange symmetry of octet color
wavefunctions of the quarks, bag-model constructions show that adding a
$J^P=1^+$ gluon to three light quarks with total quark-spin 1/2 yields
both $N$ ($I=\frac{1}{2}$) and $\Delta$ ($I=\frac{3}{2}$) hybrids with
$J^P=\sfrac{1}{2}^+$, $\sfrac{3}{2}^+$. Quark-spin 3/2 hybrids are $N$
states with $J^P=\sfrac{1}{2}^+$, $\sfrac{3}{2}^+$, and
$\sfrac{5}{2}^+$. Energies are estimated using the usual bag
Hamiltonian plus gluon kinetic energy, additional color-Coulomb
energy, and one-gluon exchange plus gluon-Compton O($\alpha_s$)
corrections. Mixings between $q^3$ and $q^3g$ states from gluon
radiation are evaluated. If the gluon self-energy is included, the
lightest $N$ hybrid state has $J^P=\frac{1}{2}^+$ and a mass between
that of the Roper resonance and the next observed $J^P=\sfrac{1}{2}^+$
state, $N(1710)$. A second $J^P=\frac{1}{2}^+$ $N$ hybrid and a
$J^P=\frac{3}{2}^+$ $N$ hybrid are expected to be 250 MeV heavier,
with the $\Delta$ hybrid states heavier still. A similar mass estimate
of about 1500 MeV for the lightest hybrid is attained in the QCD sum
rules calculation of \Ref{QCDsr}.

For this reason there has been considerable interest in the presence
or absence of light hybrid states in the $P_{11}$ and other
positive-parity partial waves in $\pi N$ scattering.  Interestingly,
quark potential models which assume a $q^3$ structure for the Roper
resonance~\cite{seeCI} predict an energy which is roughly 100 MeV too
high, and the same is true of the $\Delta(1600)$, the lightest radial
recurrence of the ground state $J^P=\frac{3}{2}^+$
$\Delta(1232)$. Furthermore, models of the electromagnetic couplings
of baryons have difficulty accommodating the substantial Roper
resonance photocoupling extracted from pion photoproduction
data~\cite{RoperPC}. Evidence for two resonances near 1440 MeV in the
$P_{11}$ partial wave in $\pi N$ scattering was cited~\cite{twopoles},
which would indicate the presence of more states in this energy region
than required by the $q^3$ model, but this has been interpreted as due
to complications in the structure of the $P_{11}$ partial wave in this
region, and not an additional physical state~\cite{CutkoskyP11}.

This letter will show that when the glue in a baryon is given the
flux-tube structure expected~\cite{paton85} from an expansion around
the strong-coupling limit of the Hamiltonian formulation of lattice
QCD (HLGT), that the lightest hybrid baryons have similar good
quantum numbers, but substantially higher energies
and different internal structure than predicted using bag models. This
structure of the glue, where the gluon degrees of freedom collectively
condense into flux-tubes, is very different from the constituent-gluon
picture of the bag model and large-$N_c$ constructions. The basis of
this model is the assumption that the dynamics relevant to the
structure of hybrids is that of confinement. At large interquark
separations, there is evidence that the flux-tube model hybrid
potential is consistent with that evaluated from lattice
QCD~\cite{paton85,adia,morningstar97}, whereas the adiabatic bag model
does not reproduce the lattice results
there~\cite{morningstar97bag}. It has also been shown~\cite{swanson98}
that a constituent-gluon model is not able to reproduce these lattice
results~\cite{morningstar97}.

The model used here to describe the glue is the nonrelativistic
flux-tube model of \Ref{paton85,barnes95}, coupled with the adiabatic
approximation, where the quarks do not move in response to the motion
of the glue (apart from moving as a rigid body in order to maintain
the center-of-mass position). Although exact only in the heavy-quark
limit, this approximation has been shown to be good for light-quark
mesons in the flux-tube model~\cite{merlin85}. In the strong coupling
limit of HLGT, flux lines (strings) with energy proportional to their
length play the role of the glue, which are modeled here by equal mass
beads with a linear potential between nearest neighbors. The total
mass of all of the beads is given by the energy in the flux lines,
which is fixed by the string tension.

Perturbations to the strong-coupling limit are provided by the
plaquette operator, which moves the flux lines, and so the beads,
perpendicular to their rest positions. Global color gauge invariance
requires that the three flux lines which emanate from the quarks must
meet at a junction, which is also modeled by a bead. Lattice QCD
studies~\cite{bali} support confinement from a Y-shaped linear
string rather than pairwise linear strings. As a single plaquette
operator cannot move the junction and leave the Y-string in its ground
state, the junction bead is given a mass higher than those on the
strings.

The ground state energy of this configuration of beads representing
the Y-string for definite quark positions ${\bf r}_i$ defines an
adiabatic potential $V_B({\bf r}_1,{\bf r}_2,{\bf r}_3)$ for the
quarks, which consists of the string energy $b\sum_i l_i$, where $b$
is the string tension and $l_i$ is the magnitude of the vector ${\bf
l}_i$ from the equilibrium junction position to the position of quark
$i$, plus the zero-point energy of the beads. The energy of the first
excited state defines a new adiabatic potential $V_H({\bf r}_1,{\bf
r}_2,{\bf r}_3)$. The simplest model incorporating the essential
degrees of freedom has one bead per string, plus a junction bead. The
mass of a bead on string $i$ is taken to be proportional to $b\,l_i$,
and the junction bead is given a higher mass. This model has nine
degrees of freedom: three string bead transverse positions ($\xi_i$)
within, and three ($z_i$) out of the plane of the quarks, and three
junction positions, in ($x$, $y$) and out ($z$) of this plane.

A nonrelativistic Hamiltonian is constructed in the small oscillations
approximation, and has the form
\begin{eqnarray}
H_{\rm string}&=&T_{\rm j}(\dot{\bf r})+
T_{\rm b}(\dot{\xi}_i,\dot{z}_i)+T_{\rm bb}+T_{\rm bj}\\ \nn
&&+V_{\rm j}({\bf r})
+V_{\rm b}(\xi_i,z_i),
\end{eqnarray}
where in the potential the string-bead (b) and junction-bead (j)
coordinates decouple, but there are terms $T_{\rm bb}$ and $T_{\rm bj}$
in the kinetic energy which couple these motions. This
Hamiltonian is corrected for the center-of-mass (c.m.) motion due to
the string motion, by allowing the quarks to move rigidly to maintain
the c.m. position. This gives effective masses similar to reduced
masses to the string and junction beads, which depend on the quark
masses.

Diagonalization of the resulting coupled-oscillator Hamiltonian for
a wide variety of quark configurations shows that neglect of the
coupling terms $T_{\rm bb}$ and $T_{\rm bj}$ does not significantly
affect the lowest string energies. Examination of the eigenfunction
for the lowest energy mode shows that it is always predominantly
in-plane junction motion. This is illustrated for three quark
positions (specified by the lengths $l_i$ of the three strings) in
\Tab{Tab:INI}.
\begin{table}
\begin{center}
\caption{Lowest three non-interacting (NI) and exact frequencies in
GeV for three quark configurations, for $m_j=m_{\rm quark}=0.33$ GeV,
and $b=0.18$ GeV$^2$.}
   \begin{tabular}{lcccccc}
     $l_i$(fm) & $E_1$(NI) & $E_1$ & $E_2$(NI) & $E_2$ 
        & $E_3$(NI) & $E_3$ \\
    \hline
     0.5, 0.5, 0.5 & 0.614 & 0.607 & 0.614 & 0.607 & 0.869 & 0.828   \\
     0.5, 0.5, 0.1 & 0.623 & 0.616 & 1.069 & 0.985 & 1.069 & 1.005   \\
     0.5, 1.0, 0.1 & 0.520 & 0.483 & 0.544 & 0.534 & 0.544 & 0.590   \\
   \end{tabular}
\label{Tab:INI}
\end{center}
\end{table}

Accordingly, the string ground state and first excited state adiabatic
surfaces may be found by allowing the junction to move, while the
strings connecting the junction to the quarks follow without
excitation. This gives the junction an effective mass which in the
limit of a large number of beads becomes
\beq 
M_{\rm eff}=b\sum_i l_i \left[\frac{1}{3} -\frac{b\sum_i
l_i}{4\sum_i (bl_i+M_i)} \right], 
\eeq
where the second term is the center of mass correction with quark
masses $M_i$. The potential is the string tension times the length of
the lines connecting the displaced junction to the quarks, so that
the junction Hamiltonian is
\begin{equation}
H_{\rm flux}=\frac{1}{2}M_{\rm eff}\dot{\bf r}^2+b\sum_{i=1}^3
\vert {\bf l}_i-{\bf r}\vert.
\label{jHam}
\end{equation}
The adiabatic surfaces are found numerically {\it via} a variational
calculation. This is made necessary by a singularity in the small
oscillations expansion for quark configurations where the triangle
made by joining them contains an angle $\theta_i$ of at least
120$^{\rm o}$, so that the equilibrium position of the junction is such
that the corresponding distance $l_i$ between the junction and the
quark is zero. This variational calculation agrees with the analytic
small oscillations results when the $l_i$ are all large, but shows the
small oscillations approximation to be poor for some other
configurations. It was shown in \Ref{barnes95} that when the small
oscillations approximation is removed, as has been done here, hybrid
meson masses go down, and when the adiabatic approximation is removed,
as has been done partially here, hybrid meson masses go up. This same
behavior is seen here.

As
$T_{\rm j}(\dot{\bf r})+V_{\rm j}({\bf r})$
is even under $z\to -z$, one of the three first excited modes of the
junction always involves motion along $\etahat_z=\hat{z}$. Analytic
results in the small oscillations approximation show that the
frequency for motion along $\etahat_z$ is always higher than those in
the plane of the quarks. Trial wavefunctions for the ground and first
excited states are taken to be the anisotropic harmonic oscillator
wavefunctions
\begin{eqnarray}
\Psi_B({\bf r})&&=
{(\alpha_+\alpha_-\alpha_z)^{1\over 2}\over \pi^{3\over 4}}
\nn \\
&&{\rm exp}\left\{-\left[(\alpha_+\etahat_+\cdot{\bf r})^2
                +(\alpha_-\etahat_-\cdot{\bf r})^2
                +(\alpha_z z)^2\right]/2\right\}  \nn\\
\Psi_H({\bf r})&&=\sqrt{2}\alpha_-\etahat_-\cdot{\bf r}
                \Psi_B({\bf r}),
\label{wvfns}
\end{eqnarray}
with variational parameters $\alpha_-$, $\alpha_+$, $\alpha_z$, and
$\theta$, the latter giving the direction $\etahat_-$ in the plane of
the quarks of the lowest-energy oscillation relative to a body-fixed
axis in that plane. For every configuration of the quarks, specified
by the magnitudes of the Jacobi coordinates $\rho$, $\lambda$, and the
cosine of the angle $\theta_{\rho\lambda}$ between them, the ground
($V_B$) and first excited state ($V_{H}$) string energies are
independently minimized. 

In order to compare to the relativized model calculation of baryon
masses of \Ref{CI}, hybrid baryon masses are found by allowing the
quarks to move in a confining potential given by the linear potential
$b\sum_i l_i$ of \Ref{CI}, plus $V_{H}-V_B$. The Coulomb potential
from one-gluon exchange is assumed the same in conventional and hybrid
baryons, and spin-dependent terms are neglected. Quark wavefunctions
are expanded in a large oscillator basis, and energies for baryons
(linear confinement) and hybrids (linear plus $V_{H}-V_B$ confinement)
composed of light quarks are evaluated. When added to the
spin-averaged mass of the $N$ and $\Delta$ which is 1085 MeV, hybrids
with quark orbital angular momenta $L_q=0,1,2$ have masses 1980, 2340
and 2620 MeV respectively. Hyperfine (contact plus tensor)
interactions split the $N$ hybrids down and the $\Delta$ hybrids up by
similar amounts, so that the $N$ hybrid mass becomes 1870 MeV. The
model error on this mass is estimated to be less than $\pm 100$
MeV. This lightest $(L_q=0$) hybrid level is substantially higher than
the roughly 1.5 GeV estimated from bag model and QCD sum rules
calculations.

This calculation also shows that the size of the quark core of
hybrid baryons is roughly 20\% larger for hybrid baryons than for
conventional baryons.

The hyperfine interaction derived from the Coulomb interaction between
the quarks has the same sign in conventional and hybrid baryons, so
that the hybrids with quark-spin-$\frac{3}{2}$ will be heavier than
those with quark-spin-$\frac{1}{2}$. Given the increased size of
the hybrid states, the quark-spin-$\frac{1}{2}$ hybrids are not
expected to split as far from the spin averaged level as in the usual
baryons.

For every set of quark positions ${\bf r}_i$ the potential in which
the junction moves is anisotropic, which means that the solutions of
the Schr\"odinger equation for the junction motion do not have
definite angular momentum. However, in the absence of the adiabatic
approximation the combined wavefunction of the quark and junction
motions must be a state of good angular momentum. Although it is
possible to project states of good angular momentum out of the
combined junction and quark motion states, this is technically
difficult and will not be reported on here. Instead, an intuitive
argument is given to justify the expectation that the total orbital
angular momentum of the lowest-lying hybrid baryons ($H$) is unity.

The $H$ hybrid wavefunction in \Eq{wvfns} is proportional to
$\hat{\lpmb{\eta}}_-\cdot{\bf r}$, and since $\hat{\lpmb{\eta}}_-$
lies in the plane of the quarks, it is proportional to a linear
combination of $Y_{11}(\hat{\bf r})$ and $Y_{1-1}(\hat{\bf r})$, with
the junction position ${\bf r}$ defined relative to a (body) $z$-axis
perpendicular to the quark plane. When the quarks are in their lowest
energy $L_q=0$ state in the hybrid confining potential, it can be
shown that the total angular momentum projection with respect to fixed
axes is $\pm 1$ as for the body axes, so that the total orbital angular
momentum must be at least $L=1$. This $L=1$ configuration is expected
to be dominant in the $H$ hybrids.

The junction Hamiltonian in \Eq{jHam} is invariant under the inversion
of coordinates, ${\bf r}_i\to-{\bf r}_i$ which implies ${\bf
l}_i\to-{\bf l}_i$, and ${\bf r}\to-{\bf r}$, so that the flux
wavefunction must be a state of good parity. Since $\etahat_-$ is a
vector in the plane of the quarks it can be written as a linear
combination of the ${\bf l}_i$, with coefficients which are functions
of the parity-invariant lengths $l_i$. It follows that $\etahat_-$ is
parity-odd, and that the hybrid-baryon wavefunction from \Eq{wvfns}
has even parity.

Permutations $P_{ij}$ of the quark labels exchange the quark positions
${\bf r}_i \leftrightarrow {\bf r}_j$, and so ${\bf l}_i
\leftrightarrow {\bf l_j}$. Since the junction Hamiltonian in
\Eq{jHam} is symmetric under such a relabeling, then $H_{\rm flux}$
and $P_{ij}$ must commute. This implies that $\Psi$ and $P_{ij}\Psi$,
where $\Psi$ is an eigenfunction of $H_{\rm flux}$ with energy $V({\bf
r}_1,{\bf r}_2,{\bf r}_3)$, must be degenerate. Since the baryon $B$
and hybrid $H$ have different energies, $P_{ij}\Psi$ must be a
multiple of $\Psi$. As the only one-dimensional representations of the
permutation group are either totally symmetric (S) or totally
antisymmetric (A), it follows that for the hybrid $\Psi$ is either S
(hybrids $H^S$) or A (hybrids $H^A$) under quark label exchange. With
the above flux Hamiltonian, the logical choice is an
exchange-symmetric baryon trial wavefunction $\Psi_B$.

The color structure of baryons and hybrid baryons is motivated by the
strong-coupling limit of lattice QCD, where the quarks are triplet
sources of color joined by strings connected to a totally
antisymmetric junction, so that the color wavefunctions of both kinds
of state are totally antisymmetric under quark-label exchange. The
Pauli principle therefore requires that the ground state symmetric
hybrids $H^S$ have combined flavor-spin wavefunctions for the quarks
which are totally symmetric, since these will have symmetric quark
orbital wavefunctions. This implies that flavor-symmetric $\Delta$
states must have $S=\frac{3}{2}$, and that flavor-mixed-symmetry $N$
states have $S=\frac{1}{2}$. The $H^A$ hybrids require totally
antisymmetric quark flavor-spin wavefunctions, which are impossible
for $\Delta$ states, and possible only for $N$ states with
$S=\frac{1}{2}$.
\begin{table}
\begin{center}
\begin{tabular}{c|l|c|l}
Hybrid Baryon          &  $L$ & $S$ & $(N,\Delta)^{2S+1}J^P$ \\
\hline
$H^S$ &  1 & $\frac{1}{2},\frac{3}{2}$  &  
$N^2 {\frac{1}{2}}^+, \; N^2 {\frac{3}{2}}^+, \; 
\Delta^4 {\frac{1}{2}}^+, \; \Delta^4 {\frac{3}{2}}^+, \; 
\Delta^4 {\frac{5}{2}}^+$\\
$H^A$ &  1 & $\frac{1}{2}$  &  $N^2 {\frac{1}{2}}^+, \; 
N^2 {\frac{3}{2}}^+$\\
\hline
{\rm bag model} & 1 & $\frac{1}{2},\frac{3}{2}$ &  
$\Delta^2 {\frac{1}{2}}^+, \; \Delta^2 {\frac{3}{2}}^+, \; 
N^4 {\frac{1}{2}}^+, \; N^4 {\frac{3}{2}}^+, \; 
N^4 {\frac{5}{2}}^+$\\
&  1 & $\frac{1}{2}$  &  $N^2 {\frac{1}{2}}^+, \; 
N^2 {\frac{3}{2}}^+$\\
\end{tabular}
\end{center}
\caption{\small Quantum numbers of low--lying hybrid baryons for the
adiabatic surface $H$, degenerate in the absence of spin-dependent forces.} 
\label{configs}
\end{table}
When these flavor-spin configurations are put together with the
expected dominant $L^P=1^+\otimes 0^+$ combined flux-quark orbital
wavefunction for the $H$ hybrids, the result is the set of
configurations shown in \Tab{configs}. Although both the present model
and the bag model~\cite{bag} predict the presence of seven low-lying
hybrid baryons, only the states $N^2{\frac{1}{2}}^+$ and
$N^2{\frac{3}{2}}^+$ have the same flavor, quark spin $S$, total
angular momentum, and parity as low-lying hybrids in the bag
model~\cite{bag}. Flavor assignments for the $H^S$ hybrids are
reversed in the bag model, due to the different (mixed) exchange
symmetry of the quark color wavefunctions required by the addition of
a single colored gluon. These differences emphasize the collective
nature of the gluonic excitation in the flux-tube model. With the
addition of spin-dependent forces, the lowest-lying hybrid states are
predicted to be two pairs of quark-spin-$\frac{1}{2}$ nucleon states
in the $P_{11}$ and $P_{13}$ partial waves, with more massive
quark-spin-$\frac{3}{2}$ $\Delta$ states. The central conclusion of
this letter is that the lightest of these states are at 1870$\pm 100$
MeV, considerably higher than previous mass estimates from the bag
model~\cite{bag} and QCD sum rules~\cite{QCDsr} of about 1500 MeV.

If hybrid baryons obey suppression of certain decay modes similar to
the decay selection rules for hybrid mesons~\cite{hybridstrong}, they
may be distinguishable from conventional baryons in the same mass
range on the basis of their strong decays; in addition, their
electromagnetic couplings are expected to be
distinctive~\cite{Roper}. There are $q^3$ baryon states predicted by
quark potential models to be present in these partial waves at these
energies which are missing from the analyses. A signal for the
presence of hybrid baryons would be the discovery, in analyses of data
expected from the new experiments, of more states than predicted by
models which constrain the glue to be in its ground
state. Furthermore, in order to understand experiments which are
designed to find these missing $q^3$ baryons, states with excited glue
must be considered. An obvious place to look for signals for hybrid
baryons would be in electro- and photoproduction of $\rho$ and
$\omega$ in Hall B at TJNAF; there are also planned $\pi N$ scattering
experiments~\cite{E913} by the Crystal Ball collaboration at the new
D-line at BNL which will examine the final states $N\eta$,
$N\rho$ and $N\omega$ in the 2 GeV mass region.

The decays $\psi\rightarrow p\bar{p}\omega$ and $\psi\rightarrow
p\bar{p}\eta^{'}$ have been observed\cite{pdg98} with branching ratios
of roughly $10^{-3}$. Since gluonic hadron production is expected to
be enhanced above conventional hadron production in the glue--rich
decay of the $\psi$, it is possible that a partial wave analysis of
the $p\omega$ or $p\eta^{'}$ invariant masses would yield evidence for
hybrid baryons. Future work in this area at BEPC and an upgraded
$\tau$--charm factory could be of critical importance.


\section{Acknowledgments}
This work is supported by the U.S. Department of Energy
under contracts No.~DE-FG05-86ER40273 (SC), No.~W-7405-ENC-36 (PRP)
and the Florida State University Supercomputer Computations Research
Institute which is partially funded by the Department of Energy through 
contract No.~DE-FC05-85ER25000.

\end{document}